\documentclass[reprint,secnumarabic,amssymb, amsmath, nobibnotes, aps, prl]{revtex4-2}
\usepackage{lineno}
\usepackage{graphicx}
\usepackage{graphics}
\usepackage[colorlinks,
            linkcolor=blue,
            anchorcolor=blue,
            citecolor=blue
            ]{hyperref}
\usepackage{amsmath}
\usepackage{graphicx}
\usepackage{dcolumn}
\usepackage{bm}

\begin{document}

\title{Single-photon blockade in quasichiral atom-photon interaction: \\simultaneous high purity and high efficiency}

\author{Yu-Wei Lu}%
\affiliation{School of Physics and Optoelectronic Engineering, Foshan University, Foshan 528000, China}
\affiliation{School of Physics and Optoelectronics, South China University of Technology, Guangzhou 510641, China}
\author{Jing-Feng Liu}%
\affiliation{College of Electronic Engineering, South China Agricultural University, Guangzhou 510642, China.}
\author{Runhua Li}%
\affiliation{School of Physics and Optoelectronics, South China University of Technology, Guangzhou 510641, China}
\author{Yanxiong Wu}%
\email[Corresponding Author: ]{wuyanxiong@fosu.edu.cn}
\affiliation{School of Physics and Optoelectronic Engineering, Foshan University, Foshan 528000, China}
\author{Haishu Tan}%
\email[Corresponding Author: ]{tanhaishu@fosu.edu.cn}
\affiliation{School of Physics and Optoelectronic Engineering, Foshan University, Foshan 528000, China}
\author{Yongyao Li}%
\affiliation{School of Physics and Optoelectronic Engineering, Foshan University, Foshan 528000, China}
\affiliation{Guangdong-Hong Kong-Macao Joint Laboratory for Intelligent Micro-Nano Optoelectronic
Technology, Foshan University, Foshan 528000, China}



\begin{abstract}
We investigate the single-photon blockade (1PB) in the quasichiral regime of atom-photon interaction that mediates via dissipative environment, where the effective atom-photon interaction is asymmetrical but achiral. The synthetic magnetic current in the closed-loop coupling breaks down the reciprocity of atom-photon interaction, resulting in an asymmetrical and even completely unidirectional effective coupling between two selected quantum states. As an example, we couple the single-atom cavity-QED (cQED) system to a strongly dissipative auxiliary cavity. We find that in the quasichiral regime, the unconventional photon blockade (UPB) always incorporates with the conventional photon blockade (CPB) in the condition of maximum chirality. Furthermore, we show that 1PB in the quasichiral regime combines the advantages of UPB and CPB, demonstrating the perfect single-photon purity, higher efficiency, smooth time dynamics as well as lower requirement of modes coupling to achieve UPB. Our work paves the way for 1PB towards practical applications and reveals the intriguing quantum-optics phenomena in the quasichiral light-matter interaction. 
\end{abstract}


\maketitle

\section{I. Introduction}
A $n$-photon quantum light source that emits at most $n$ photons at a time, especially the single-photon source and two-photon source, has unique applications in various quantum technologies, ranging from quantum computation and information \cite{RN1,RN2,RN45} to quantum detection \cite{RN3}. The conventional scheme utilizes the anharmonicity of the discrete level structure of single atom in cavity to block the additional photon \cite{RN4}. The mechanism is analogy to the Coulomb blockade effect and thus called as the photon blockade \cite{RN5}. Another unconventional scheme is to induce the destructive interference between all possible transition pathways of a targeted n-photon state, then the $n$-1 photon blockade can occur \cite{RN6,RN7,RN8}. However, two existing mechanisms of photon blockades, CPB and UPB, both have obvious drawbacks \cite{RN9}: CPB requires the sufficiently strong atom-cavity coupling to have good performance, while the efficiency of UPB is low and its delayed two-photon correlation $g^{(2)} (\tau)$ generally manifests fast oscillations. These hinder the practical applications of photon blockades. 

\begin{figure}[t]
\includegraphics[width=8.5cm]{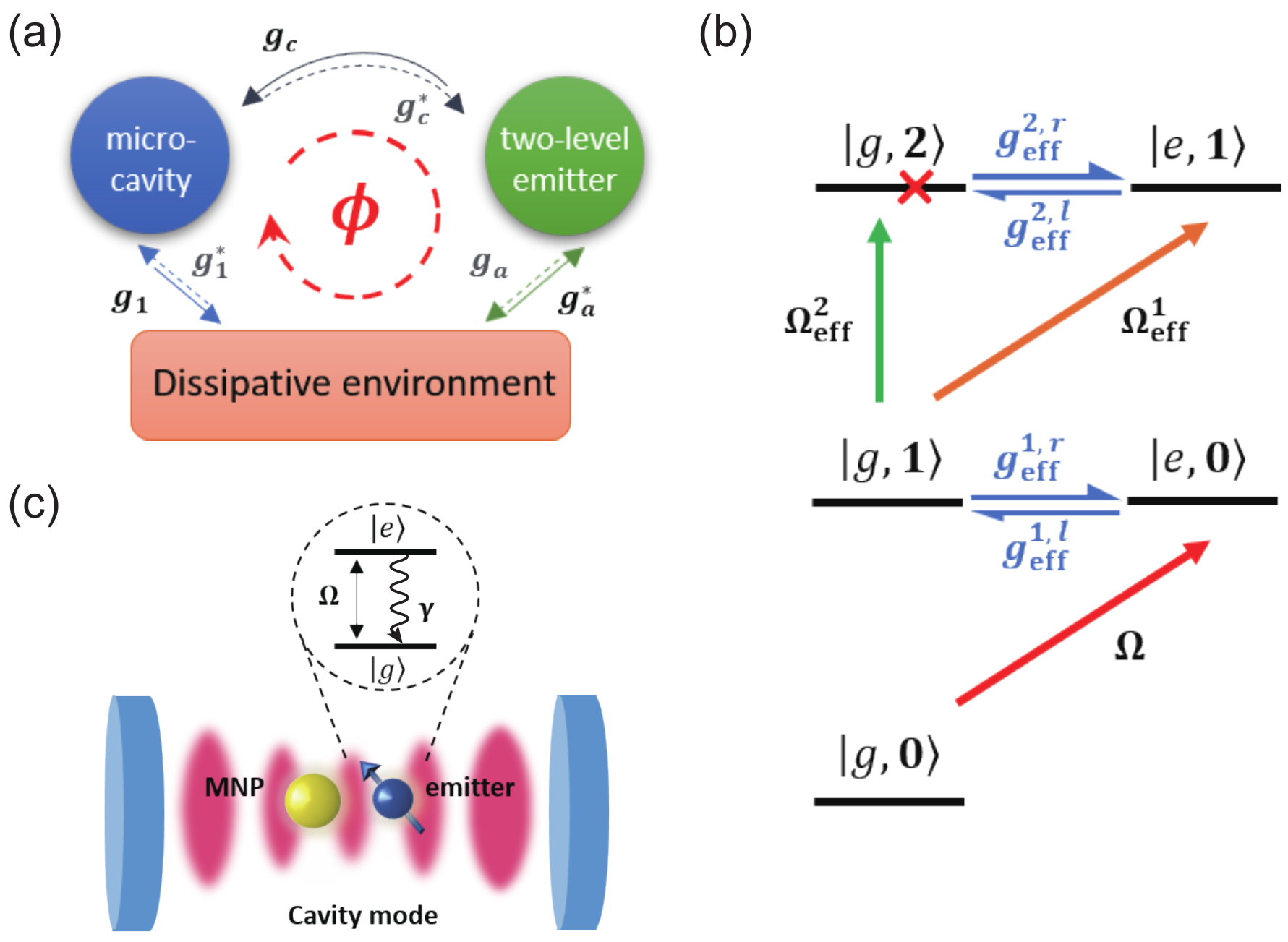}%
\caption{(a) Schematic of the interaction of a Jaynes–Cummings cQED system coupled to the dissipative environment. All couplings are reciprocal but with complex coupling strengths, resulting in an accumulated phase in the triangular loop. (b) Transition pathways of the effective atom-photon interaction in the system shown in (a) for atom-driven case with non-reciprocal coupling that can be either chiral or quasichiral. Blue arrows represent the coupling direction with strengths $g_{\rm{eff}}^{(2,r)}$, $g_{\rm{eff}}^{(2,l)}$, $g_{\rm{eff}}^{(1,r)}$ and $g_{\rm{eff}}^{(1,l)}$, standing for the cavity-to-atom coupling and the cavity-to-atom coupling in the single-excitation and two-excitation subspace, respectively. $\Omega$ is the driving strength, while $\Omega_{\rm{eff}}$ represents the effective driving. (c) A sketch of the mutually coupled QE-MNP-microcavity system with QE driven by a classical laser field. \label{fig1}}
\end{figure}

The previous studies of single-photon blockade (1PB) are in the reciprocal regime of atom-photon interaction \cite{RN8,RN44} and photon-photon interaction \cite{RN10,RN11}. The situation is completely different when the reciprocity is broken. For the single-atom cQED system, UPB cannot take place in the case of atom drive, due to the lack of the second transition pathway that is required for generating the quantum interference \cite{RN7,RN8}. The system behaviors just like a single atom with the chiral cavity-to-atom coupling, since the cavity has no effect on the atom. Furthermore, the energy level splitting is absent in the unidirectional coupling, and therefore, CPB will not occur at the chiral regime. In this work, we consider the quasichiral atom-photon interaction mediated by the dissipative environment instead of the unidirectional coupling in a cascaded quantum system \cite{RN12,RN13,RN46}, as the schematic illustrated in Fig. \ref{fig1}(a). The essential difference is that in the former case, the interaction between each component is reciprocal. In the case of closed-loop coupling, the nonzero accumulated phase of complex coupling strength, i.e., the existence of synthetic magnetic current, breaks down the reciprocity \cite{RN14,RN15,RN16,RN17,RN18}. Tuning the synthetic magnetic current can control the chirality of the effective coupling between two selected quantum states. This feature is in favor of manipulating the quantum states to obtain the desirable quantum-optics properties. For the atom-driven configuration, the dissipative environment provides an additional effective pump for cavity, see the transition pathways for effective atom-photon interaction shown in Fig. \ref{fig1}(b). The synthetic magnetic current can tune the effective pump and the atom-photon coupling, and hence change the energy levels, leading to the complex of quantum interference between different quantum states and the rich phenomena of photon blockades. 

\section{II. Physical model for quasichiral atom-photon interaction}
We construct a quantum system with similar nonreciprocity induced by the synthetic magnetic field \cite{RN16,RN19}. Fig. \ref{fig1}(c) illustrates the system we consider. An auxiliary cavity represented by the metallic nanoparticle (MNP) couples to a typical single-atom cQED system, and interacts with both the cavity mode and the quantum emitter (QE)\cite{RN20,RN21,RN22}. The interactions in the system is sketched in Fig. \ref{fig1}(a). The QE is modeled as a two-level atom, and the MNP is seen as a dissipative environment for its large decay rate. The nonzero accumulated phase in this closed-loop coupling gives rise to the chirality, which cannot be eliminate via gauge transformation like that in the Jaynes-Cummings model \cite{RN23}. 

To highlight the difference between our system and a general single-mode cavity, we consider the atom-driven case, where UPB cannot occur. In the frame rotating at driving frequency, the system Hamiltonian is written as
\begin{equation}
\begin{aligned}
H&= \Delta \omega_{c} c^{\dagger} c+\Delta \omega_{a} a^{\dagger} a+\Delta \omega_{0} \sigma_{+} \sigma_{-}+\left(g_{a} a^{\dagger} \sigma_{-}+g_{a}^{*} \sigma_{+} a\right) \\
&+\left(g_{c} c^{\dagger} \sigma_{-}+g_{c}^{*} \sigma_{+} c\right)+\left(g_{1} a^{\dagger} c+g_{1}^{*} a c^{\dagger}\right)+\Omega\left(\sigma_{+}+\sigma_{-}\right)
\end{aligned}
\end{equation}
where $c$ and $a$ are the bosonic annihilation operators for microcavity and the dipolar plasmonic mode of MNP, with resonance frequencies $\omega_c$ and $\omega_a$, respectively. $\sigma_{-}=|g\rangle\langle e|$ denotes the lowing operator of the QE, and $\omega_0$ is the atom transition frequency. $\Delta \omega_{X}=\omega_{X}-\omega_{L}$ represents the frequency detuning with respect to pump field, with $\omega_L$ being the laser frequency and indices $X=a,c,0$. $g_a=G_a e^{i\phi_a }$, $g_c=G_c e^{i\phi_c}$ and $g_1=G_1 e^{i\phi_1}$ are the complex coupling constants for atom-plasmon interaction, atom-photon interaction and plasmon-photon interaction, respectively. $\Omega$ is the Rabi frequency of pump field. In the following discussion we only consider the atom-driven case unless special note. Taking the dissipations into consideration, the system is described by the quantum master equation
\begin{equation}
\dot{\rho}=i[\rho, H]-\kappa_{a} \mathcal{L}(a)-\kappa_{c} \mathcal{L}(c)-\gamma \mathcal{L}(\gamma)
\end{equation}
where $\mathcal{L}(\hat{o})=\hat{o}^{\dagger} \hat{o} \rho / 2+\rho \hat{o}^{\dagger} \hat{o} / 2-\hat{o} \rho \hat{o}^{\dagger}$ is the Lindblad dissipator. $\kappa_a$, $\kappa_c$ and $\gamma$ are the decay rates of the modes $a$, $c$ and the QE, respectively. In the case of weak pump, we can drop the quantum jump term $\hat{o} \rho \hat{o}^{\dagger}$ in the Lindblad dissipator to obtain the effective Hamiltonian
\begin{equation}
\begin{aligned}
H&= \Delta_{c} c^{\dagger} c+\Delta_{a} a^{\dagger} a+\Delta_{0} \sigma_{+} \sigma_{-}+\left(g_{a} a^{\dagger} \sigma_{-}+g_{a}^{*} \sigma_{+} a\right) \\
&+\left(g_{c} c^{\dagger} \sigma_{-}+g_{c}^{*} \sigma_{+} c\right)+\left(g_{1} a^{\dagger} c+g_{1}^{*} a c^{\dagger}\right)+\Omega\left(\sigma_{+}+\sigma_{-}\right)
\end{aligned}\label{eq1}
\end{equation}
where $\Delta_{X}=\left(\omega_{X}-\omega_{L}\right)-i \kappa_{X} / 2$ with $\kappa_0 = \gamma$. Now the effective Hamiltonian is non-Hermitian and can describe both coherent and dissipative dynamics. 

\subsection{A. Single-photon intensity and zero-delayed two-photon correlation}
To calculate the zero-delayed two-photon correlation $g^{(2)} (0)$, we truncate the state space by two-excitation manifold and as a result, the time-dependent wave function $|\psi(t)\rangle$ is expressed as
\begin{equation}
|\psi(t)\rangle=\sum_{n+m+p \leq 2, n=0,1} C_{n m p}|n\rangle_{e}|m\rangle_{a}|p\rangle_{c}\label{eq2}
\end{equation}
where $C_{nmp}$ is the coefficient of quantum state $|n\rangle_{e}|m\rangle_{a}|p\rangle_{c}$. $m$ and $p$ stand for the photon number in the MNP and the microcavity, respectively. $n=0$ and 1 represent the atom in the ground state and the excited state, respectively. From the Schr\"odinger equation, we can obtain the equations of motion for coefficients 
\begin{subequations}
\begin{widetext}
\begin{align}
&i \dot{C}_{001}=\Delta_{c} C_{001}+g_{c} C_{100}+g_{1}^{*} C_{010} \label{eq3} \\
&i \dot{C}_{010}=\Delta_{a} C_{010}+g_{a} C_{100}+g_{1} C_{001} \\
&i \dot{C}_{100}=\Delta_{0} C_{100}+g_{a}^{*} C_{010}+g_{c}^{*} C_{001}+\Omega \label{eq5} \\
&i \dot{C}_{110}=\left(\Delta_{a}+\Delta_{0}\right) C_{110}+\sqrt{2} g_{a}^{*} C_{020}+g_{c}^{*} C_{011}+g_{1} C_{101}+\Omega C_{010} \\
&i \dot{C}_{101}=\left(\Delta_{c}+\Delta_{0}\right) C_{101}+g_{a}^{*} C_{011}+\sqrt{2} g_{c}^{*} C_{002}+g_{1}^{*} C_{110}+\Omega C_{001} \\
&i \dot{C}_{011}=\left(\Delta_{a}+\Delta_{c}\right) C_{011}+g_{a} C_{101}+g_{c} C_{110}+\sqrt{2} g_{1} C_{002}+\sqrt{2} g_{1}^{*} C_{020} \\
&i \dot{C}_{002}=2 \Delta_{c} C_{002}+\sqrt{2} g_{c} C_{101}+\sqrt{2} g_{1}^{*} C_{011} \\
&i \dot{C}_{020}=2 \Delta_{a} C_{020}+\sqrt{2} g_{a} C_{110}+\sqrt{2} g_{1} C_{011} \\
\end{align}
\end{widetext}
\end{subequations}
We are interested in the photon statistics of microcavity, where the photon can be easy to output and transport via the coupling to waveguide. By setting $i \dot{C}_{n m p}=0$, we can obtain the steady-state solutions for the coefficients corresponding to the photon states of microcavity
\begin{widetext}
\begin{gather}
C_{001}=\frac{\Omega\left(G_{1} G_{a} e^{-i \phi}-\Delta_{a} G_{c}\right)}{D_{1}} e^{i \phi_{c}} \label{eq6} \\ 
C_{002}=D_{1} \frac{C_{001}^{2}\left[\Delta_{a}^{2}+\left(\Delta_{a} \Delta_{c}+\left|g_{1}\right|^{2}\right)+\left(\Delta_{a} \Delta_{0}-\left|g_{a}\right|^{2}\right)+\left(\Delta_{c} \Delta_{0}-\left|g_{c}\right|^{2}\right)\right]-g_{1}^{*} C_{010}\left(\Delta_{c} C_{001}-g_{c} C_{100}\right)}{\sqrt{2} D_{2}}
\end{gather}
\end{widetext}
with coefficients 
\begin{gather}
C_{010}=\frac{\Omega\left(G_{1} G_{c} e^{i \phi}-\Delta_{c} G_{a}\right)}{D_{1}} e^{i \phi_{a}} \\
C_{100}=\frac{\Omega\left(\Delta_{a} \Delta_{c}-\left|g_{1}\right|^{2}\right)}{D_{1}} 
\end{gather}
and determinants
\begin{equation}
D_{1}=\left|\begin{array}{ccc}
\Delta_{c} & g_{1}^{*} & g_{c} \\
g_{1} & \Delta_{a} & g_{a} \\
g_{c}^{*} & g_{a}^{*} & \Delta_{0}
\end{array}\right|
\end{equation}
\begin{equation}
D_{2}=\left|\begin{array}{ccccc}
\Delta_{a}+\Delta_{0} & g_{1} & g_{c}^{*} & 0 & \sqrt{2} g_{a}^{*} \\
g_{1}^{*} & \Delta_{c}+\Delta_{0} & g_{a}^{*} & \sqrt{2} g_{c}^{*} & 0 \\
g_{c} & g_{a} & \Delta_{a}+\Delta_{c} & \sqrt{2} g_{1} & \sqrt{2} g_{1}^{*} \\
0 & \sqrt{2} g_{c} & \sqrt{2} g_{1}^{*} & 2 \Delta_{c} & 0 \\
\sqrt{2} g_{a} & 0 & \sqrt{2} g_{1} & 0 & 2 \Delta_{a}\label{eq11}
\end{array}\right|
\end{equation}
where the relative phase $\phi=\phi_{a}-\phi_{1}-\phi_{c}$. The averaged photon number of microcavity $I_c \approx\left|C_{001} \right|^2$ and corresponding the zero-delayed two-photon correlation $g_c^{(2)} (0)\approx\left|\sqrt{2}C_{002} \right|^2/I_c^2$ can be analytically obtained from Eqs.(\ref{eq6})-(\ref{eq11}). One can verify that both $I_c$ and $g_c^{(2)}(0)$ are only determined by the relative phase $\phi$ instead of the absolutely phase of the coupling strengths. 

\subsection{B. Effective atom-photon interaction and chirality}
From Eq. (\ref{eq6}), we can see that when the conditions $\phi=3\pi/2$ and $G_a=-G_c \kappa_a/2G_1$ are satisfied, $C_{001}$ is exactly equal to zero at $\Delta \omega_a=0$, i.e., there is no photon in the microcavity. This nontrivial phenomenon is due to the chiral cavity-to-atom coupling in the case of atom drive, and cannot be realized by the quantum interference of transition pathways in the reciprocal regime with $\phi=0$. To get a better understanding and gain physical insight, we adiabatically eliminate the mode $a$ in Eqs. (\ref{eq3}) to (\ref{eq5}) to obtain the following non-Hermitian effective Hamiltonian
\begin{equation}\label{eq17}
H_{\mathrm{eff}}^{a c} \approx\left[\begin{array}{cc}
\omega_{c}-\frac{i \kappa_{c}}{2}\left(1+\frac{4 G_{1}^{2}}{\kappa_{a} \kappa_{c}}\right) & G_{\mathrm{c}}+i \frac{2 G_{1} G_{\mathrm{a}}}{\kappa_{a}} e^{i \phi}  \\
G_{\mathrm{c}}+i \frac{2 G_{1} G_{\mathrm{a}}}{\kappa_{a}} e^{-i \phi} & \omega_{0}-\frac{i \gamma}{2}\left(1+\frac{4 G_{a}^{2}}{\kappa_{a} \gamma}\right)
\end{array}\right]
\end{equation}
where two non-diagonal elements are not conjugate due to the appearence of dissipation-mediated coupling. This leads to the loss of reciprocity and asymmetrical coupling between the microcavity and the QE. We define the chirality of this effective interaction as
\begin{equation}
C=\frac{\left|H_{12}\right|-\left|H_{21}\right|}{\left|H_{12}\right|+\left|H_{21}\right|}
\end{equation}
where $H_{12}=G_{\mathrm{c}}+i 2 G_{1} G_{\mathrm{a}} e^{i \phi} / \kappa_{a}$ and $H_{21}=G_{\mathrm{c}}+i 2 G_{1} G_{\mathrm{a}} e^{-i \phi} / \kappa_{a}$ are the non-diagonal elements in $H_{\mathrm{eff}}^{a c}$. $C=\pm 1$ is the chiral point that corresponds to the chiral atom-to-cavity/cavity-to-atom coupling in the single-excitation subspace, while $C=0$ is the common reciprocal atom-photon interaction. Except for these three special cases, other values of $C$ lie in the quasichiral regime, where the coupling between the cavity and the QE is asymmetrical but still bidirectional. Accordingly, we can define $H_{12}=0$, $H_{21}\neq0$ and $H_{21}=0$, $H_{12}\neq0$ as the chiral conditions, which yields the same results obtained from the full Hamiltonian (Eq. (\ref{eq1})), i.e., by letting $C_{001}=0$. Actually, from Eq. (\ref{eq6}) we can see that $C_{001}$ can be expressed as $C_{001}=\Omega H_{21}/{D_1}$ if $\Delta \omega_a=0$. 

Previous study has shown that the atom-photon quasi-bound state (qBS) can form in this cQED system with $\omega_a=\omega_c$ and $\omega_0-\omega_c=G_c (G_a/G_1-G_1/G_a ) \cos(\phi)$, obtained from Eq. (\ref{eq17}) by finding the condition of a purely real eigenvalue, which is shown to yield the highest single-photon efficiency \cite{RN24}. In order to achieve best 1PB performance apart from the maximum chirality ($\phi\neq3\pi/2$ and $\phi\neq\pi/2$), in the following study we will apply the aforementioned condition of atom-photon qBS. However, it can be seen that at the maximum chirality, the condition of qBS is trivial since $\rm{cos}(\phi)=0$, which means the resonant coupling between the QE, the microcavity and the plasmonic antenna. It is the case we mainly focus on in this work. 

\section{III. Single-photon blockade in quasichiral regime}
\begin{figure}[t]
\includegraphics[width=8.5cm]{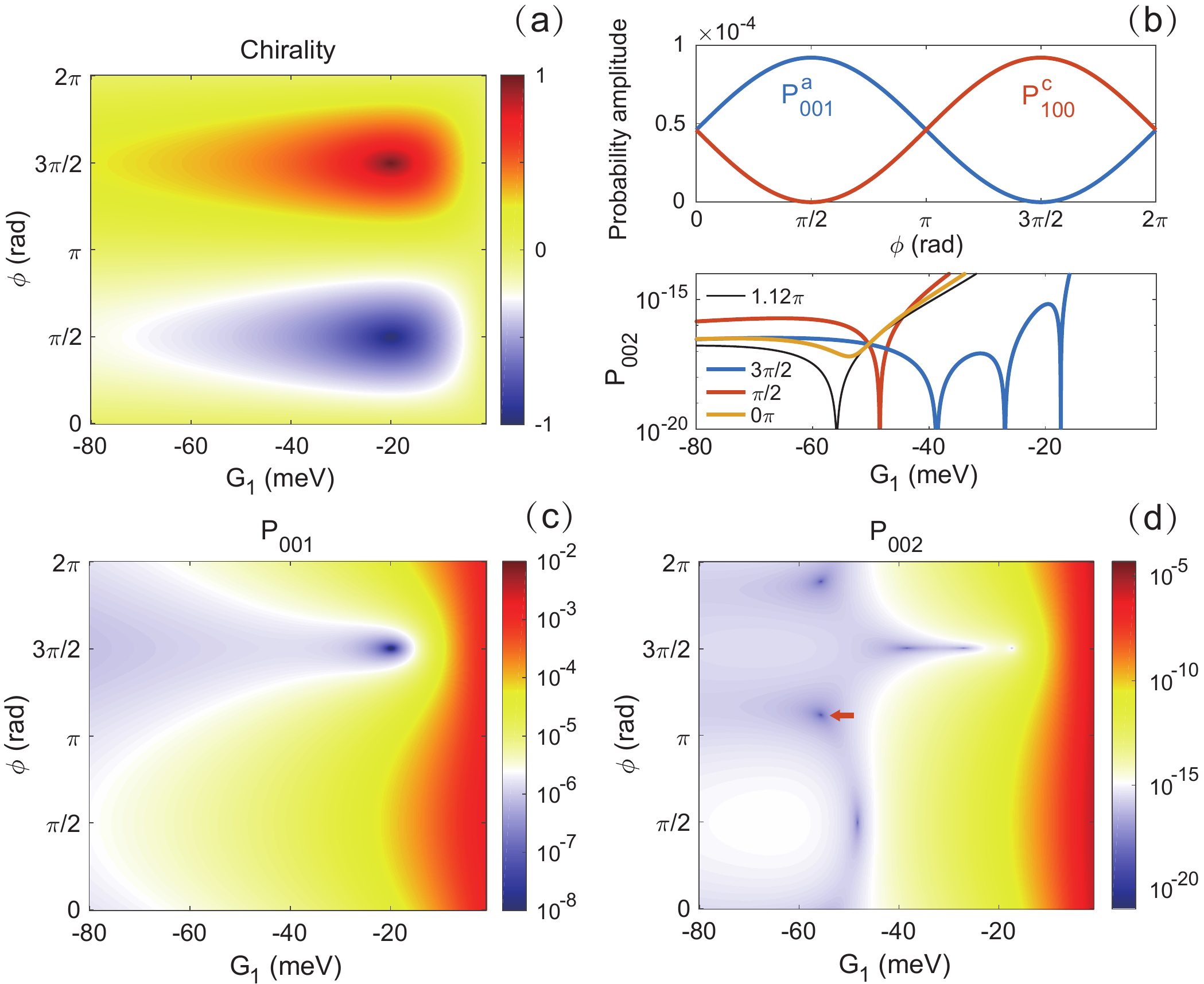}%
\caption{(a) The change of the chirality $C$ with respect to the magnitude of the modes coupling $G_1$ and the phase $\phi$. (b) Upper panel: the probability amplitude of one-photon state $|g, 0,1\rangle$ $P_{001}$ as the function of $\phi$ for atom drive (blue line) and cavity drive (red line) at chiral point with $G_1=-20$meV. The superscript $c$ and $a$ mean the atom drive and the cavity drive, respectively. Lower panel: the probability amplitude of two-photon state $|g, 0,2\rangle$ $P_{002}$ as the function of $G_1$ for variou $\phi$ under the atom drive. (c) $P_{001}$ versus $G_1$ and $\phi$. (d) $P_{002}$ versus $G_1$ and $\phi$. The red arrow indicates the location of 1PB for $\phi=1.12\pi$ (black line in the upper panel of (b)). Other parameters are $\kappa_c=0.5$meV, $\kappa_1=0.1$eV, $\gamma=0.1$meV, $G_c=0.2$meV, and $\Omega=0.1\gamma$.  \label{fig2}}
\end{figure}

For hybrid plasmonic-photonic cavity, the coupling strengths of QE with the microcavity ($G_c$) and the plasmonic antenna ($G_a$) can be tuned by varying the QE location, while the coupling strength between the microcavity and the plasmonic antenna ($G_1$) is hard to tune once the structure has been fabricated. Therefore, we first analyze the chirality of atom-photon interaction with respect to $G_1$. In Fig. \ref{fig2}(a), we plot the chirality as the function of $G_1$ and $\phi$ with $G_a=-G_c \kappa_a/2G_1^s$, where $G_1^s=-20$meV, and this value of $G_1$ can be achieved in the recently reported hybrid cavities \cite{RN15,RN41}. As a result, the chiral point $C=\pm1$ will appear at $G_1=G_1^s$, and we can see that $C$ reaches 1 at $\phi=3\pi/2$ and $-1$ at $\phi=\pi/2$. The chirality degrades faster away from the chiral point for $G_1>-20$meV than $G_1<-20$meV. For $G_1<-20$meV, the effective atom-photon interaction is mainly contributed by the plasmon-photon coupling and thus is dissipative, while for $G_1>-20$meV the coherent atom-photon coupling is dominant. At the chiral points, the atom-photon interaction is purely coherent and unidirectional. We can expect that the system operated at the chiral points, for example, in the case of chiral atom-to-cavity coupling, driving the cavity will not excite the atom while driving the atom will produce higher averaged photon number of cavity compared to a reciprocal system. Fig. \ref{fig2}(b) demonstrates such circumstance, where we plot the probability $P_{100}^{c}=\left|C_{100}\right|^{2}$ for cavity drive and $P_{001}^{a}=\left|C_{001}\right|^{2}$ for atom drive as the function of $\phi$. The alternate variation of $P_{100}^c$ and $P_{001}^a$ clearly indicates the change of reciprocity in the atom-photon interaction. Note that only at $C=\pm 1$, i.e., with $G_1=G_1^s$ and $\phi=\pi/2$ or $\phi=3\pi/2$, the atom-photon interaction is chiral according to Eq. (\ref{eq17}). For any $G_1\neq-20$meV, the maximum chirality is still achieved at $\phi=\pi/2$ and $\phi=3\pi/2$, while in this case the atom-photon interaction is asymmetrical but achiral, and thus is in the quasichiral regime.

Different from the unidirectional interaction in a cascaded single-atom cQED system, which prevents the coupling of opposite direction at all the subspace with different excitations, here the chiral interaction is effective, and hence is merely selectively achieved at one state while the coupling between other states remain achiral, as shown in Fig. \ref{fig2}(c) and (d). We can see that $P_{001}=\left|C_{001}\right|^2$ and $P_{002}=\left|C_{002}\right|^2$ reach zero with different parameters of $G_1$ and $\phi$. This feature enables us to realize the desirable $n$-photon induced tunneling and blockade, which is impossible for the corresponding cascaded system.

\begin{figure}[t]
\includegraphics[width=9cm]{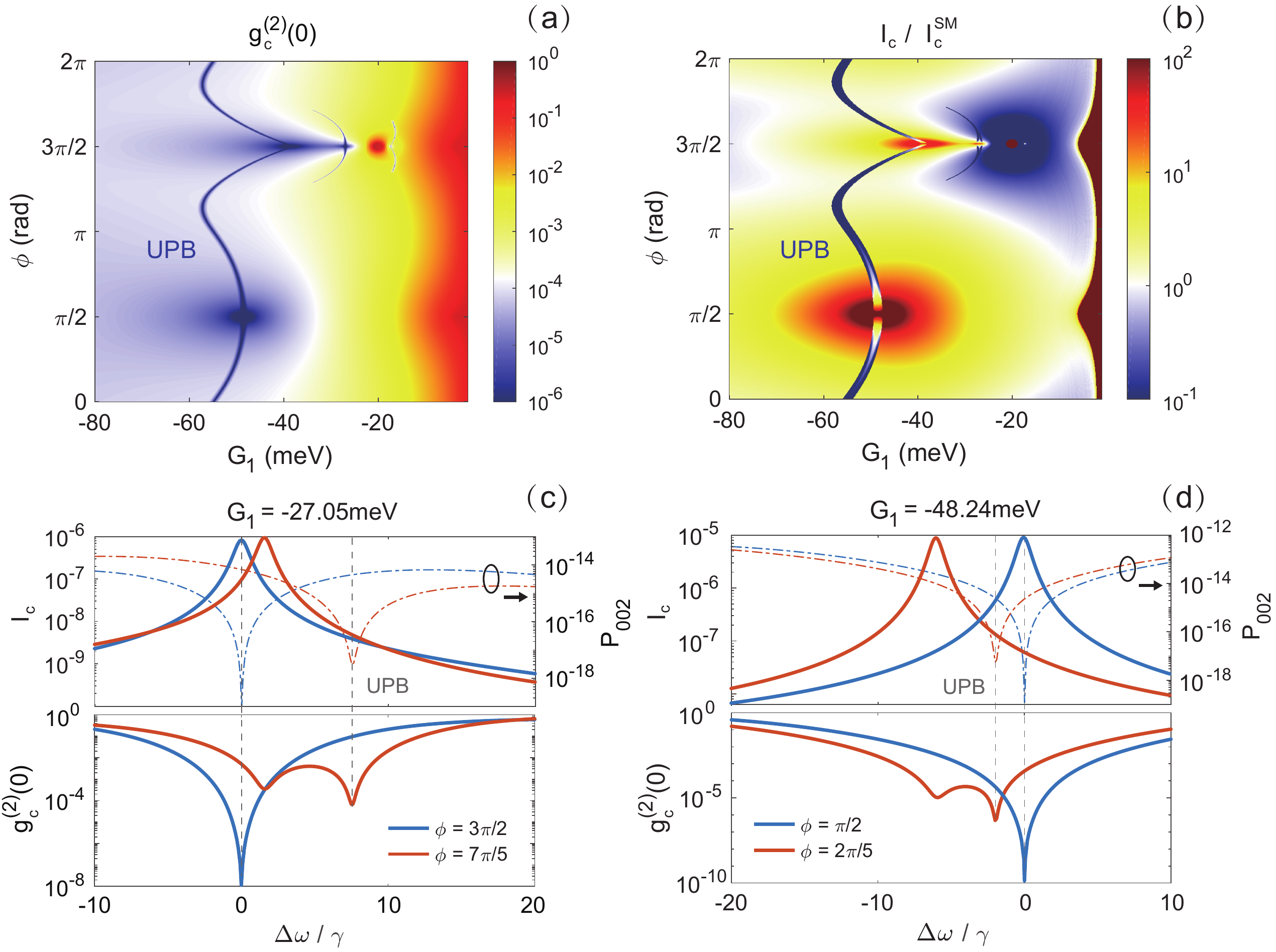}%
\caption{(a) The zero-delayed two-photon correlation $g_c^{(2)} (0)$ of microcavity as the function of $G_1$ and $\phi$. (b) The enhancement of the averaged photon number of microcavity $I_c$ as the function of $G_1$ and $\phi$, compared to the single-atom cQED system without the MNP when achieving the same $g_c^{(2)} (0)$, denoting as $I_c^{SM}$. (c) Upper panel: $I_c$ (solid lines) and $P_{002}$ (dashed dotted lines) versus laser detuning $\Delta\omega$ for different $\phi$ with $G_1=-27.05$meV. Lower panel: corresponding $g_c^{(2)} (0)$. (d) is the same as (c) but for $G_1=-48.24$meV and different $\phi$. Other parameters are the same as Fig. \ref{fig2}.  \label{fig3}}
\end{figure}

In this study, we focus on 1PB that occurs at the zero points of $P_{002}$. In Fig. \ref{fig2}(d), we can see that there are two zero points at the line $\phi=3\pi/2$ and one at $\phi=\pi/2$, see also the lower panel of Fig. \ref{fig2}(b). Notably, there are also two zero points of $P_{002}$ lie outside the lines of $\phi=3\pi/2$ and $\phi=\pi/2$, but are symmetrically distributed with respect to the line $\phi=3\pi/2$. By comparing the $P_{002}$ at $\phi=0$ where a dip at $G_1\approx-57$meV can be seen, we are aware that the zero points located at $G_1<-30$meV is the UPB that inherits from the incomplete quantum interference in the reciprocal regime. This can be clearly seen and identified from the corresponding $g_c^{(2)} (0)$ shown in Fig. \ref{fig3}(a), where the UPB dip continuously varies when $\phi$ changes from 0 to $2\pi$. However, such a modes coupling is much larger than that of reported structures \cite{RN15,RN41}, where the maximum $G_1$ is around 35meV. Therefore, UPB is hard to observe in realistic hybrid cavity under the reciprocal light-matter interaction. The remaining zero point of $P_{002}$ at $\phi=3\pi/2$ are unique in the quasichiral regime, but also belongs to UPB. On the other hand, in Fig. \ref{fig3}(a) we can also see that the photon induced tunneling (PIT) occurs around the chiral point $\phi=3\pi/2$ and $G_1=-20$meV. However, different from the PIT in the reciprocal regime, this PIT results from the reduced probability of single-photon state $P_{001}$ instead of the increased probability of two-photon state $P_{002}$. Actually, from Fig. \ref{fig2}(d) we can see that $P_{002}$ is also reduced around the chiral point $\phi=3\pi/2$. 

\begin{figure}[t]
\includegraphics[width=8.5cm]{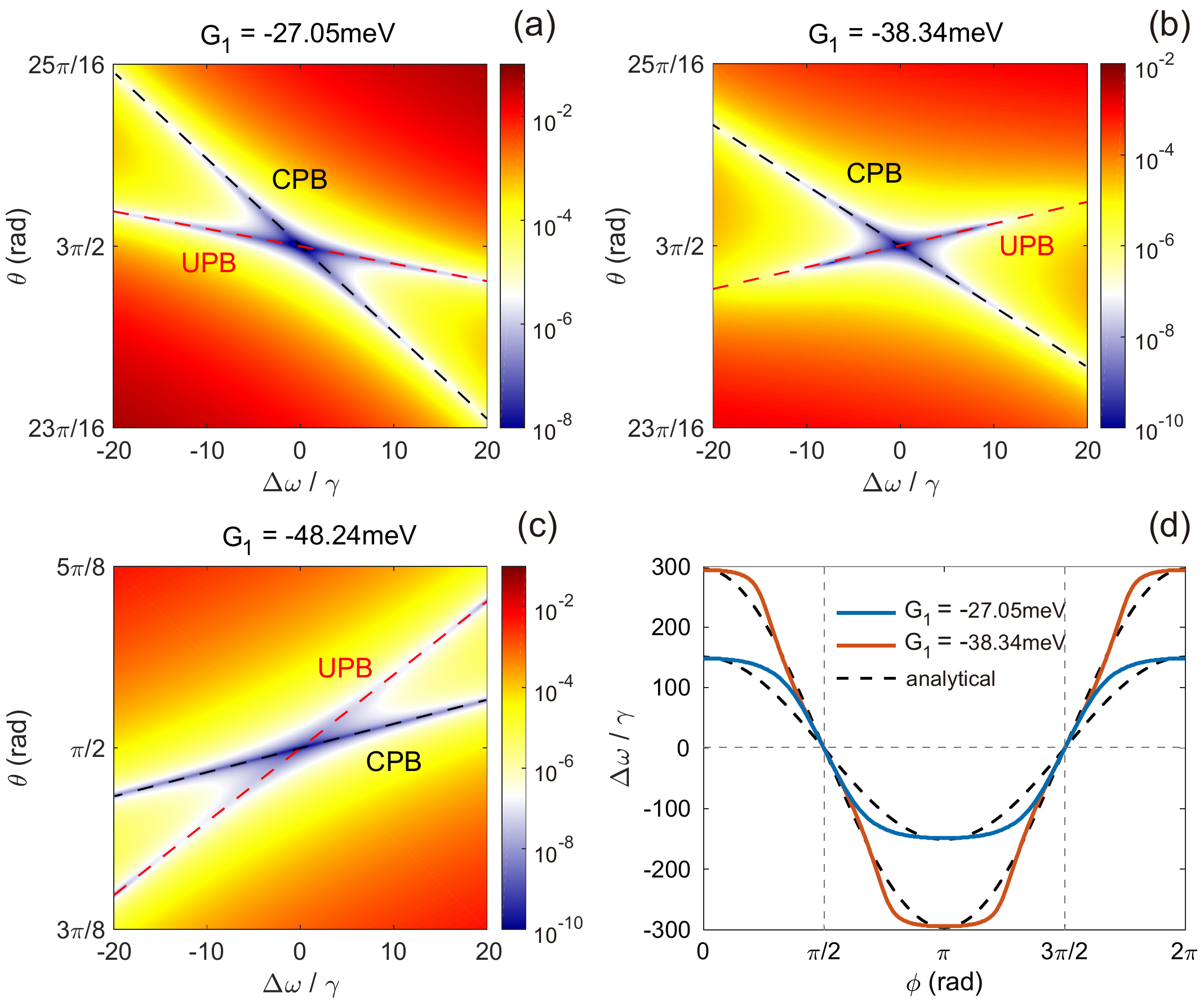}%
\caption{(a) $g_c^{(2)} (0)$ as the function of $\Delta\omega$ and $\phi$ for $G_1=-27.05$meV. The dashed black and red lines respectively trace the CPB and UPB. (b) and (c) are the same as (a), but for $G_1=-38.34$meV and $-48.24$meV, respectively. (d) The peak location of $I_c$ versus $\phi$ for different $G_1$. The dashed black lines show the result given by the analytical expression Eq. (\ref{eq14}). Dashed gray lines are provided as a guide to the eye. Other parameters are the same as Fig. \ref{fig2}.  \label{fig4}}
\end{figure}

Although both the UPB in the reciprocal and nonreciprocal regimes can achieve perfect single-photon purity, the latter has much higher efficiency (larger $I_c$). Fig. \ref{fig3}(b) shows the enhancement of the efficiency as the function of $G_1$ and $\phi$, compared to the single-atom cQED system achieving the same $g_c^{(2)} (0)$ via varying the atom-cavity detuning. We can see that in the reciprocal regime, the efficiency is enhanced at a large range of parameters, and giant enhancement ($>10^2$) is found around the chiral points, especially for $\phi=\pi/2$ where the nonreciprocity increases $I_c$. Due to $g_c^{(2)} (0)=0$ at the zero points of $C_{002}$, the enhancement factor is diverging at these points in Fig. \ref{fig2}(c). 
In Fig. \ref{fig3}(c), we plot the $I_c$ and $g_c^{(2)} (0)$ as the function of laser detuning $\Delta\omega=\Delta\omega_a$ at the chiral point $\phi=3\pi/2$ and $G_1=-27.05$meV, and the case of that away from the chiral point with $\phi=7\pi/5$ is also shown for comparison. We can see that the minimum of $g_c^{(2)} (0)$, which is also the minimum of $P_{002}$, corresponds to the maximum of $I_c$ at chiral point, while there are two dips of $g_c^{(2)} (0)$ for $\phi=7\pi/5$. One dip corresponds to the peak of $I_c$, another results from the dip of $P_{002}$. Similar phenomenon is also observed at the chiral point $\phi=\pi/2$ and $G_1=-48.24$meV, as shown in Fig. \ref{fig3}(d). We can infer that 1PB at two chiral points is the mixture of CPB and UPB, and they separate when the system deviates from the chiral point. This behavior can be clearly seen in Fig. \ref{fig4}(a)-(c), where we plot the $g_c^{(2)} (0)$ as the function of $\Delta\omega$ and $\phi$ for three chiral points with different $G_1$ and $\phi$. The dashed black lines trace the dip corresponding to CPB in $g_c^{(2)} (0)$. It shows that the CPB dip always locates at $\Delta\omega=0$ for $\phi=3\pi/2$ and $\phi=\pi/2$, but shifts toward the opposite direction as $\phi$ increases. 

To explain this feature, we find the approximate expression for the real part of the eigenvalues with smallest decaying and denote as $\omega_{qBS}$, which corresponds to CPB, from either the Hermitian (Eq. (\ref{eq1})) or the non-Hermitian effective Hamiltonian (Eq. (\ref{eq17})) 
\begin{equation}\label{eq14}
\omega_{q B S} \approx-G_{\mathrm{c}} \frac{G_{1}}{G_{\mathrm{a}}} \cos (\phi)
\end{equation}
where we can see that $\omega_{qBS}=0$ if $\rm{cos}(\phi)=0$. Fig. \ref{fig4}(d) plots Eq. (\ref{eq14}) for different $G_1$, which accords well with the exact locations of CPB dip in the whole range of $\phi$. Physically, the shift of this atom-like eigenenergy level results from the co-existence of coherent and dissipative components in the effective atom-photon interaction for $\rm{cos}(\phi)\neq0$, see the non-diagonal elements of Eq. (\ref{eq17}). While with $\rm{cos}(\phi)=0$, from the non-diagonal elements of Eq. (\ref{eq17}) we can see that the $\pi/2$ phase shift of the coupling pathway mediated by the dissipative environment is compensated and becomes in-phase with the direct one. Therefore, in this case the effective coupling strength of atom-photon interaction is purely real and coherent. 

\begin{figure}[b]
\includegraphics[width=8.5cm]{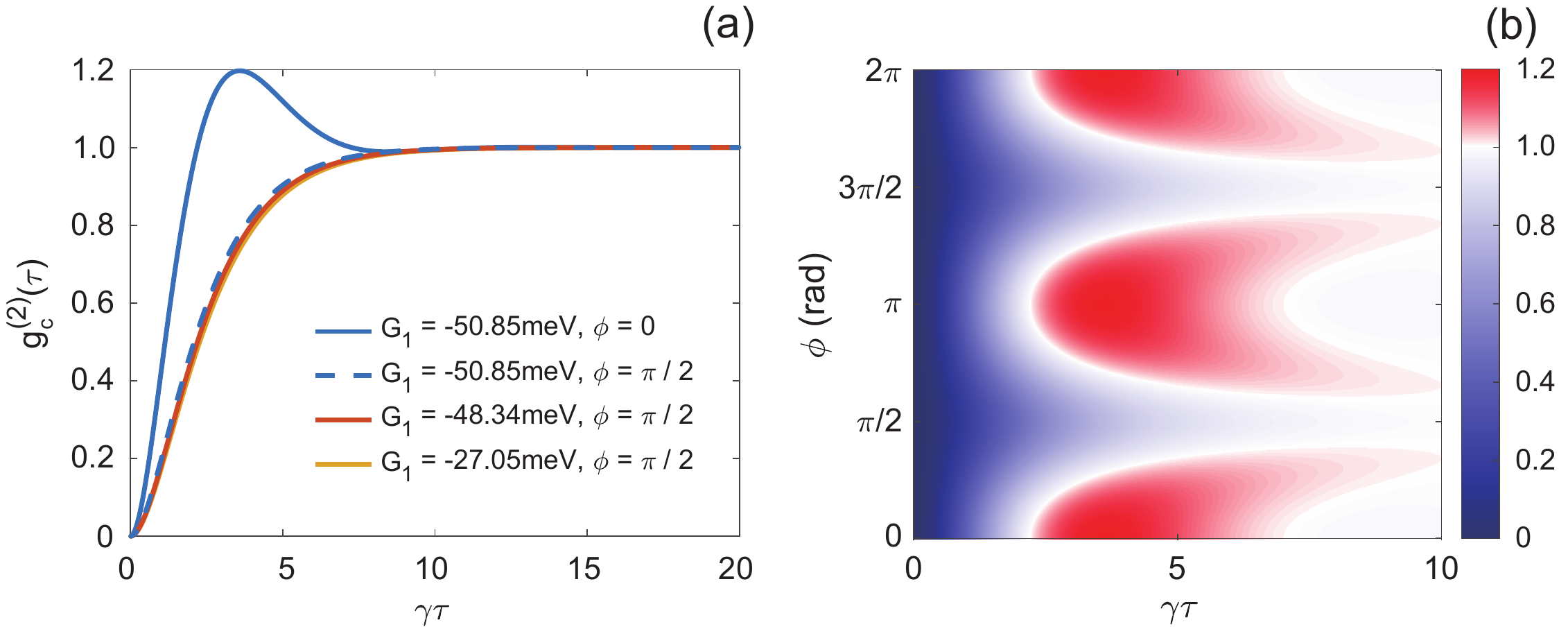}%
\caption{(a) The delayed two-photon correlation $g_c^{(2)} (\tau)$ for various $G_1$ and $\phi$. (b) $g_c^{(2)} (\tau)$ versus $\phi$ for $G_1=-50.85$meV. Other parameters are the same as Fig. \ref{fig2}. $g_c^{(2)} (\tau)$ is numerically calculated using QuTip \cite{RN25}.  \label{fig5}}
\end{figure}

The combination of CPB and UPB at the chiral points makes 1PB inherit the respective advantages of both kinds of photon blockades. Except the perfect single-photon purity, the delayed two-photon correlation $g_c^{(2)} (\tau)$ gets rid of the fast oscillations in UPB due to the modes splitting. For example, we plot the 1PB in reciprocal and nonreciprocal regime for $G_1=-50.85$meV at $\Delta\omega=0$ in Fig. \ref{fig5}(a), where we can see that the 1PB demonstrates a smooth time dynamics when approaching $\cos (\phi)=0$. Fig. \ref{fig5}(b) compares the $g_c^{(2)} (\tau)$ with $\phi=0$ and $\pi/2$. There is an obvious time oscillation of $g_c^{(2)} (\tau)$ in the reciprocal regime ($\phi=0$), which reaches the maximum 1.2 before damping to 1. While $g_c^{(2)} (\tau)$ with $\phi=\pi/2$ smoothly evolves to 1. It is because the incorporation of UPB with CPB at chiral points eliminates the oscillations in $g_c^{(2)} (\tau)$, as we discuss above. Fig. \ref{fig5}(b) also shows that the rising edge of $g_c^{(2)} (\tau)$ for various $G_1$ with $\phi=\pi/2$ only manifests slightly difference, for it is mainly determined by the decay rate of MNP. Therefore, our system does not require strong mode coupling to obtain better 1PB performance.

In the above discussion we mainly focus on the effect of the coupling strength between the microcavity and the plasmonic mode ($G_1$) and the induced phase ($\phi$) on the intensity $I_c$ and the second-order correlation function $g_c^{(2)} (0)$. We further investigate the influence of the system decays and the coupling strength between the microcavity and the QE, which turn out to be crucial in the photon blockade effect \cite{RN9}.

\begin{figure}[t]
\includegraphics[width=8.5cm]{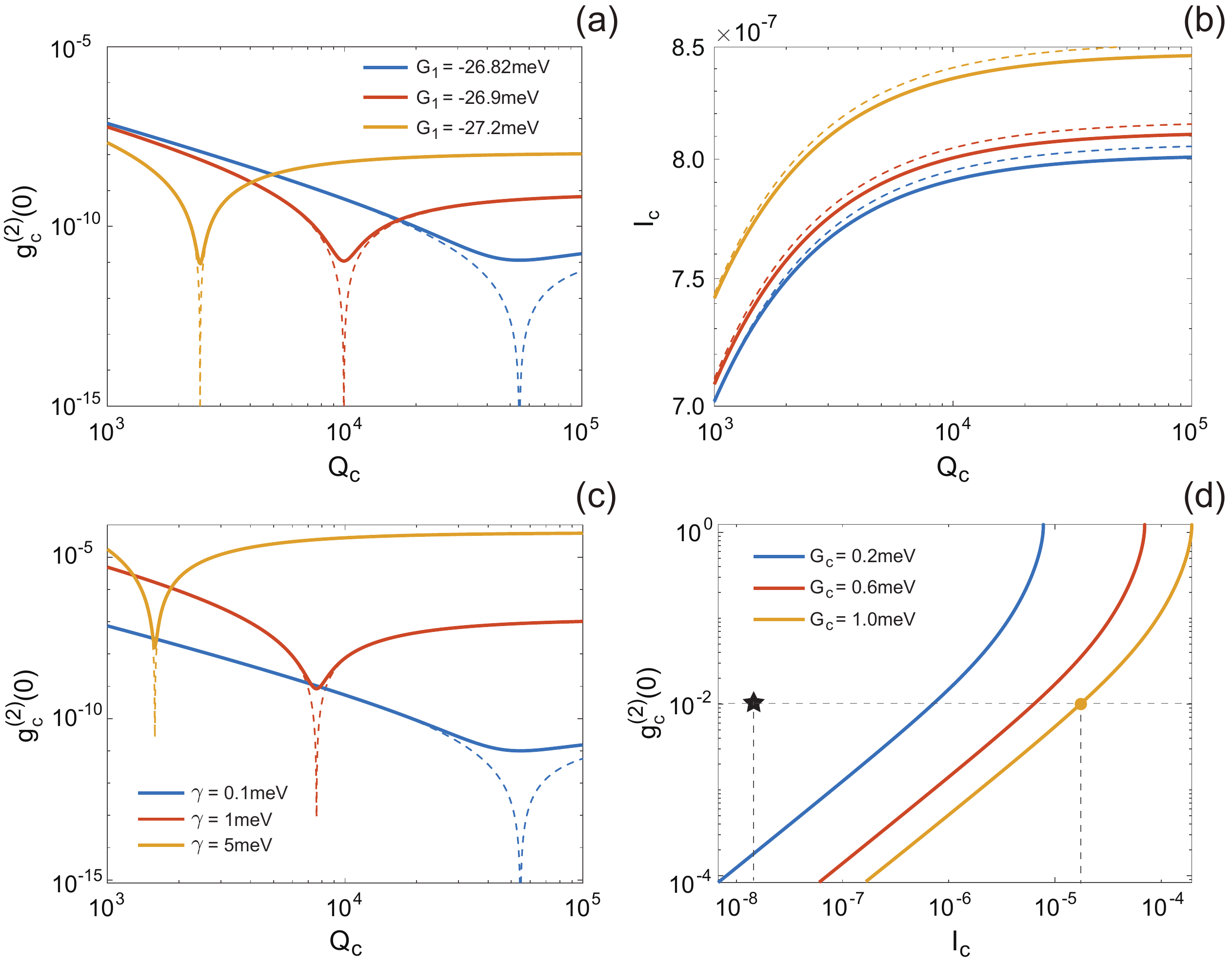}%
\caption{(a) and (b) $g_c^{(2)} (0)$ and $I_c$ versus the quality factor $Q_c$ for various $G_1$ with $\phi=3\pi/2$, respectively. (c) $g_c^{(2)} (0)$ as the function of $Q_c$ for various $\gamma$ with $G_1=-27.05$meV and $\phi=3\pi/2$. (d) $g_c^{(2)} (0)-I_c$ relation for various $G_c$ with $\gamma=20$meV, $G_1=-25.5$meV and $\phi=3\pi/2$. The black star labels the $g_c^{(2)} (0)$ and $I_c$ for a bare microcavity with $\gamma=20$meV. In the evaluation of $Q_c$ we set $\omega_c=2$eV. The solid lines are the numerical results calculated using QuTip \cite{RN25}, while the dashed lines show the results of analytical expressions. The parameters not mentioned are the same as Fig. \ref{fig2}. \label{fig6}}
\end{figure}
Fig. \ref{fig6}(a) plots the $g_c^{(2)} (0)$ as the function of the quality factor of microcavity, $Q_c=\omega_c/\kappa_c$, ranging from $10^3$ to $10^5$. This range of quality factor covers the common microcavities, such as nanobeam cavity \cite{RN26}, photonic crystal L3 cavity \cite{RN27}, and the whispering-gallery-mode (WGM) microdisk \cite{RN28} and microring \cite{RN29}. We can see that the optimal $Q_c$ for 1PB is very sensitive to the value of $G_1$. For example, as $G_1$ changes 0.3meV from -26.9meV to -27.2meV, the optimal $Q_c$ shifts from 1600 to $10^4$. On the other hand, we can see that even outside the region of 1PB, $g_c^{(2)} (0)$ is still smaller than $10^{-7}$, which implies that the deviation of optimal $Q_c$ will not lead to the observable degradation of single-photon purity in practical application. Furthermore, the efficiency of 1PB is insensitive to the variation of $Q_c$, as shown in Fig. \ref{fig6}(b). It shows that the larger $Q_c$ benefits for higher efficiency, but the difference is less then 15\% when $Q_c$ is improved by two orders of magnitude. This feature indicates that 1PB in the quasichiral light-matter interaction enables fabricating robust single-photon source and can be implemented in various microcavity structures. 

Fig. \ref{fig6}(c) shows the $g_c^{(2)} (0)$ versus $Q_c$ for various QE decays, where we can see that the large $\gamma$ seriously reduces the single-photon purity. The minimum $g_c^{(2)} (0)$ increases by about an order of magnitude when $\gamma$ changes from 1meV to 5meV, and the optimal $Q_c$ decreases as well. However, it also suggests that the room-temperature 1PB is achievable with suitable system parameters. Fig. \ref{fig6}(d) plots the purity-efficiency curves of 1PB for various $G_c$ with $\gamma=20$meV, which is the typical value of quantum dots in room temperature \cite{RN30}. It shows that the curve moves horizontally as $G_c$ increases. For $G_c=1$meV, the efficiency of 1PB demonstrates an giant improvement of three orders of magnitude compared to a bare microcavity.

The decay rate of realistic plasmonic antenna $\kappa_a$ is typically 90-200meV, according to different shapes and materials \cite{RN31,RN32}. The variation range of $\kappa_a$ is much smaller than the decays of microcavity and QE, and hence the change of $\kappa_a$ has less impact on $g_c^{(2)} (0)$ and $I_c$.

\section{IV. Physical realization}
Now that we have shown the advantageous 1PB performance in the quasichiral regime of atom-photon interaction, we briefly discuss the experimental realizations and the generation and control of the required synthetic magnetic current. The coupling strengths are generally complex number in realistic hybrid cavities \cite{RN17}, which can be evaluated via the quasinormal modes (QNM) method \cite{RN15,RN33} or the two-mode mapping \cite{RN24,RN34}. The synthetic magnetic current can be tuned by varying the position of either the QE or the plasmonic antenna. In addition, the existence of synthetic magnetic current between the hybrid cavity and the environment has been observed in experiment and extracted via coupled mode theory \cite{RN18}. Therefore, it is also possible to tune the synthetic magnetic current via environmental backaction. Another promising platform for realizing 1PB in quasichiral regime is the single QE in WGM microdisk, where the field distributions of the clockwise (CW) and counterclockwise (CCW) modes have an azimuthal spatial dependence. The synthetic magnetic current between the QE and the two modes can be tuned by varying the QE position relative to the azimuthal distribution of two modes \cite{RN35,RN47}, or alternatively by coupling the microdisk to a waveguide with mirror at one end \cite{RN36}.

\section{V. Conclusion}
In conclusion, we demonstrate the possibility of realizing the fascinating 1PB with perfect single-photon purity, higher efficiency and smooth time dynamics in the quasichiral regime of the effective atom-photon interaction. The key is to introduce the synthetic magnetic current via dissipation-mediated interaction. We highlight that the 1PB we show here is not a special case when CPB and UPB occur at the same time, like that in the reciprocal regime. In our model, the CPB and UPB always coalesce at $\phi=\pi/2$ and $3\pi/2$ with suitable $G_1$, due to the pure coherent effective interaction resulting from the loss of reciprocity. This is essentially distinguishing from the accidental combination of CPB and UPB in the reciprocal regime. Furthermore, one can verify that the UPB condition in the reciprocal regime requires $G_1<-50$meV, which is hard to reach in the existing nanophotonic structures \cite{RN15,RN22,RN37,RN41}. While in the quasichiral regime, we show that the UPB can achieve with lower modes coupling, such as $G_1\approx-27$meV. The physical realization of such quantum system is not limited by the hybrid cavities in nanophotonics we study here, but can also be implemented in waveguide QED \cite{RN34,RN38,RN42,RN43} and superconducting circuit \cite{RN39,RN40}. We believe our work is not only innovational in designing 
photon blockades but also significant in the practical applications.

\section{Acknowledgement}
This work is supported by the National Natural Science Foundation of China (Grant No. 11874112, 11874438, 62005044) and the International Joint Laboratory for Micro-Nano Manufacturing and Measurement Technologies (Grant No. 2020B1212030010).

\bibliography{q1PB.bib}

\end{document}